\documentclass[doublecol]{epl2}
\usepackage{amssymb,amsmath,graphicx,setspace,color,rotating,subfigure,url}
\title{Superfamily classification of nonstationary time series based on DFA scaling exponents}

\author{Chuang Liu\inst{1,2} \and Wei-Xing Zhou\inst{1,2,3,4}\footnote{e-mail: wxzhou@ecust.edu.cn}} \shortauthor{C. Liu \etal}

\institute{
  \inst{1} School of Business, East China University of Science and Technology, Shanghai 200237, China\\
  \inst{2} Engineering Research Center of Process Systems Engineering (Ministry of Education), East China University of Science and Technology,   Shanghai 200237, China \\ %
  \inst{3} School of Science, East China University of Science and Technology, Shanghai 200237, China\\
  \inst{4} Research Center for Econophysics, East China University of Science and Technology, Shanghai 200237, China
}


\abstract{The superfamily phenomenon of time series with different dynamics can be characterized by the motif rank patterns observed in the nearest-neighbor networks of the time series in phase space. However, the determinants of superfamily classification are unclear. We attack this problem by studying the influence of linear temporal correlations and multifractality using fractional Brownian motions (FBMs) and multifractal random walks (MRWs). Numerical investigations unveil that the classification of superfamily phenomenon is uniquely determined by the detrended fluctuation analysis (DFA) scaling exponent $\alpha$ of the time series. Only four motif patterns are observed in the simulated data, which are delimited by three DFA scaling exponents $\alpha \simeq 0.25$, $\alpha \simeq 0.35$ and $\alpha \simeq 0.45$. The validity of the result is confirmed by stock market indexes and turbulence velocity signals.}

\begin{document}

\maketitle

\section{Introduction}

In the last few decades, the understanding of the dynamics and the topological structures of complex systems has experienced extremely rapid progress. The behavior of a complex system is usually recorded and exhibited in the form of time series. New light has been shed on these studies from the perspective of complex networks \cite{Albert-Barabasi-2002-RMP,Newman-2003-SIAMR,Boccaletti-Latora-Moreno-Chavez-Hwang-2006-PR}, which considers the interactions of different constituents of a complex system though a portfolio of time series. More recently, a wealth of methods have been proposed to construct complex networks from single time series \cite{Zhang-Small-2006-PRL,Li-Wang-2006-CSB,Li-Wang-2007-PA,Zhang-Sun-Luo-Zhang-Nakamura-Small-2008-PD,Li-Yang-Komatsuzak-2008-PNAS,Xu-Zhang-Small1-2008-PNAS,Lacasa-Luque-Ballesteros-Luque-Nuno-2008-PNAS,Luque-Lacasa-Ballesteros-Luque-2009-PRE,Kostakos-2009-PA,Shirazi-Jafari-Davoudi-Peinke-Tabar-Sahimi-2009-JSM,Marwan-Donges-Zou-Donner-Kurths-2009-PLA,Donner-Zou-Donges-Marwan-Kurths-2009-XXX}, which enables us to study the dynamics of a complex system through the topological structure of the constructed network. These methods have been applied to investigate diverse natural and social systems \cite{Yang-Yang-2008-PA,Gao-Jin-2009-PRE,Gao-Jin-2009-Chaos,Ni-Jiang-Zhou-2009-PLA,Lacasa-Luque-Luque-Nuno-2009-EPL,Elsner-Jagger-Fogarty-2009-GRL,Yang-Wang-Yang-Mang-2009-PA,Liu-Zhou-Yuan-2009-XXX,Qian-Jiang-Zhou-2009-XXX}, among which the visibility graph algorithm has attracted most attention \cite{Lacasa-Luque-Ballesteros-Luque-Nuno-2008-PNAS,Luque-Lacasa-Ballesteros-Luque-2009-PRE,Ni-Jiang-Zhou-2009-PLA,Lacasa-Luque-Luque-Nuno-2009-EPL,Elsner-Jagger-Fogarty-2009-GRL,Yang-Wang-Yang-Mang-2009-PA,Liu-Zhou-Yuan-2009-XXX,Qian-Jiang-Zhou-2009-XXX}.

Many complex biological, technological and social networks share similar universal statistical properties, such as the small world characteristics \cite{Watts-Strogatz-1998-Nature,Amaral-Scala-Barthelemy-Stanley-2000-PNAS} and the scale-free distribution of node degrees \cite{Barabasi-Albert-1999-Science}. The global properties can be used to classify networks into different universality classes from a macroscopic point view \cite{Amaral-Scala-Barthelemy-Stanley-2000-PNAS}. On the other hand, complex networks with
the same global statistics may exhibit different local structure properties. At the microscopic level, the building blocks of complex networks are subgraphs or motifs \cite{Milo-ShenOrr-Itzkovitz-Kashtan-Chklovskii-Alon-2002-Science}, and the network motif patterns of occurrence can be utilized to define superfamilies of networks \cite{Milo-Itzkovitz-Kashtan-Levitt-ShenOrr-Ayzenshtat-Sheffer-Alon-2004-Science}. The superfamily phenomenon is also observed in time series, which is able to distinguish different types of dynamics in periodic, chaotic and noisy processes based on the occurrence frequency patterns of network motifs \cite{Xu-Zhang-Small1-2008-PNAS}.

In this Letter, we attempt to seek possible determinants that cause different motif rank patterns emerging in complex nearest-neighbor networks constructed from a specific time series whose increments are stationary. Numerical investigations conducted on fractional Brownian motions (FBMs) and multifractal random walks (MRWs) uncover that the superfamily phenomenon is uniquely determined by the DFA scaling exponent and the multifractality in MRWs has negligible impact. We also study the superfamily phenomenon using several stock indexes of developed and emerging markets and the velocity signals of three-dimensional fully developed turbulent flows, and find that the above numerical conclusion holds as well.

\section{Nearest neighbor networks and motifs}

The procedure to construct a nearest-neighbor network from a time series is described as follows \cite{Xu-Zhang-Small1-2008-PNAS}. The time series is embedded in a proper phase space, in which the time delay $\tau$ is calculated according to the mutual information method \cite{Fraser-Swinney-1986-PRA} and the embedding dimension $d$ should be larger than $2D+1$ to make sure that the attractors are unfolded, where $D$ is the correlation dimension of the system. Each point in the phase space is considered as a node of the network. For each node, four links are generated to its nearest neighbors that have not been connected to it. In this way, a nearest-neighbor network is constructed.

For a time series with length $\ell$, the nearest-neighbor network has $n=\ell-d\tau$ nodes and the average degree of node is 8. In this Letter, we choose $\ell=10^4$ and $d=10$, and the results presented are averaged over many realizations. Certainly, different orders of chosen nodes in constructing connections result in different networks. However, our results are independent of the node order. Since all the constructed networks have the same size, we can directly compare the motif distributions without any reference to random networks \cite{Milo-ShenOrr-Itzkovitz-Kashtan-Chklovskii-Alon-2002-Science,Milo-Itzkovitz-Kashtan-Levitt-ShenOrr-Ayzenshtat-Sheffer-Alon-2004-Science,Xu-Zhang-Small1-2008-PNAS}.

\begin{figure}[htb]
\centering
\includegraphics[width=1.3cm]{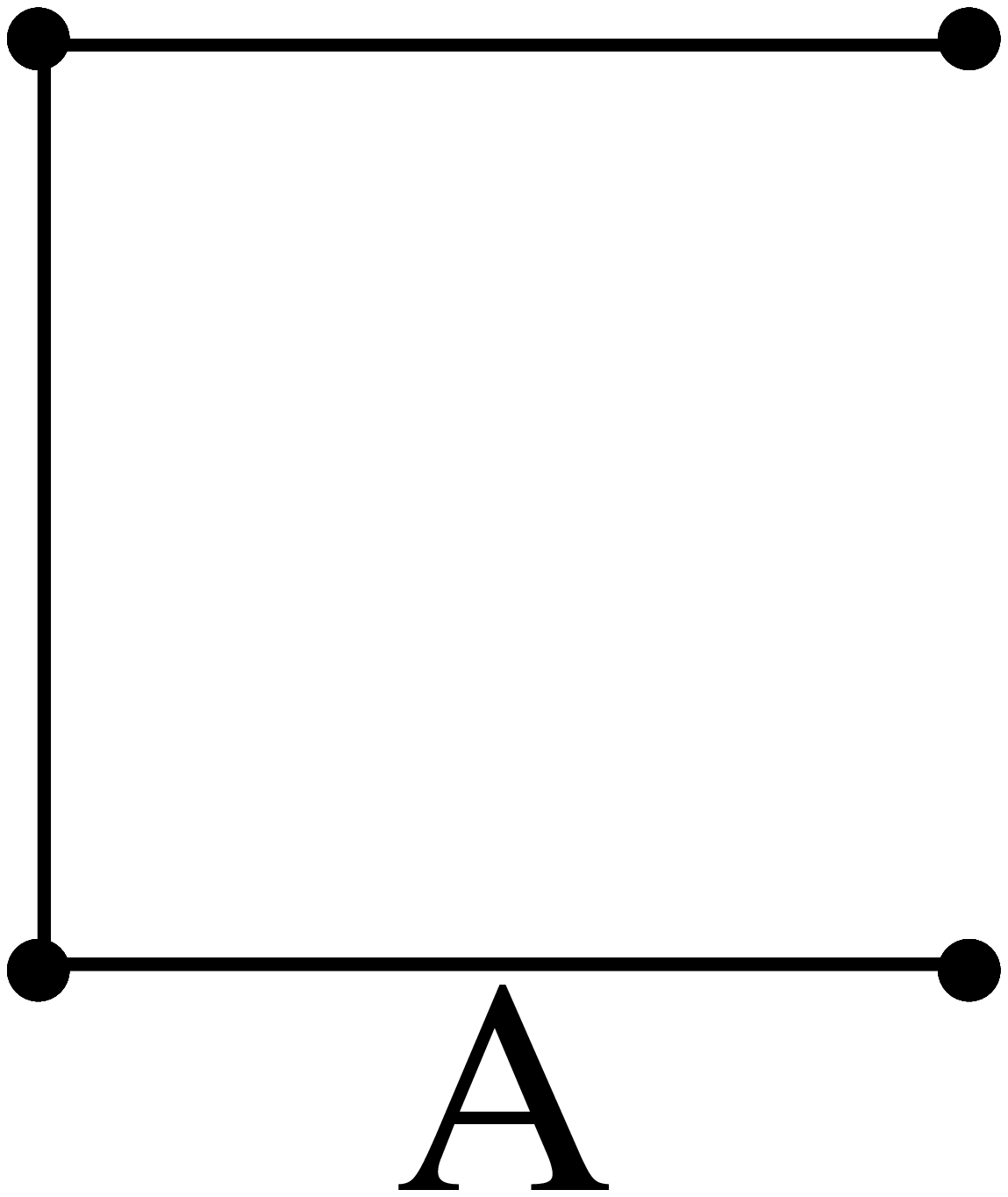}
\includegraphics[width=1.3cm]{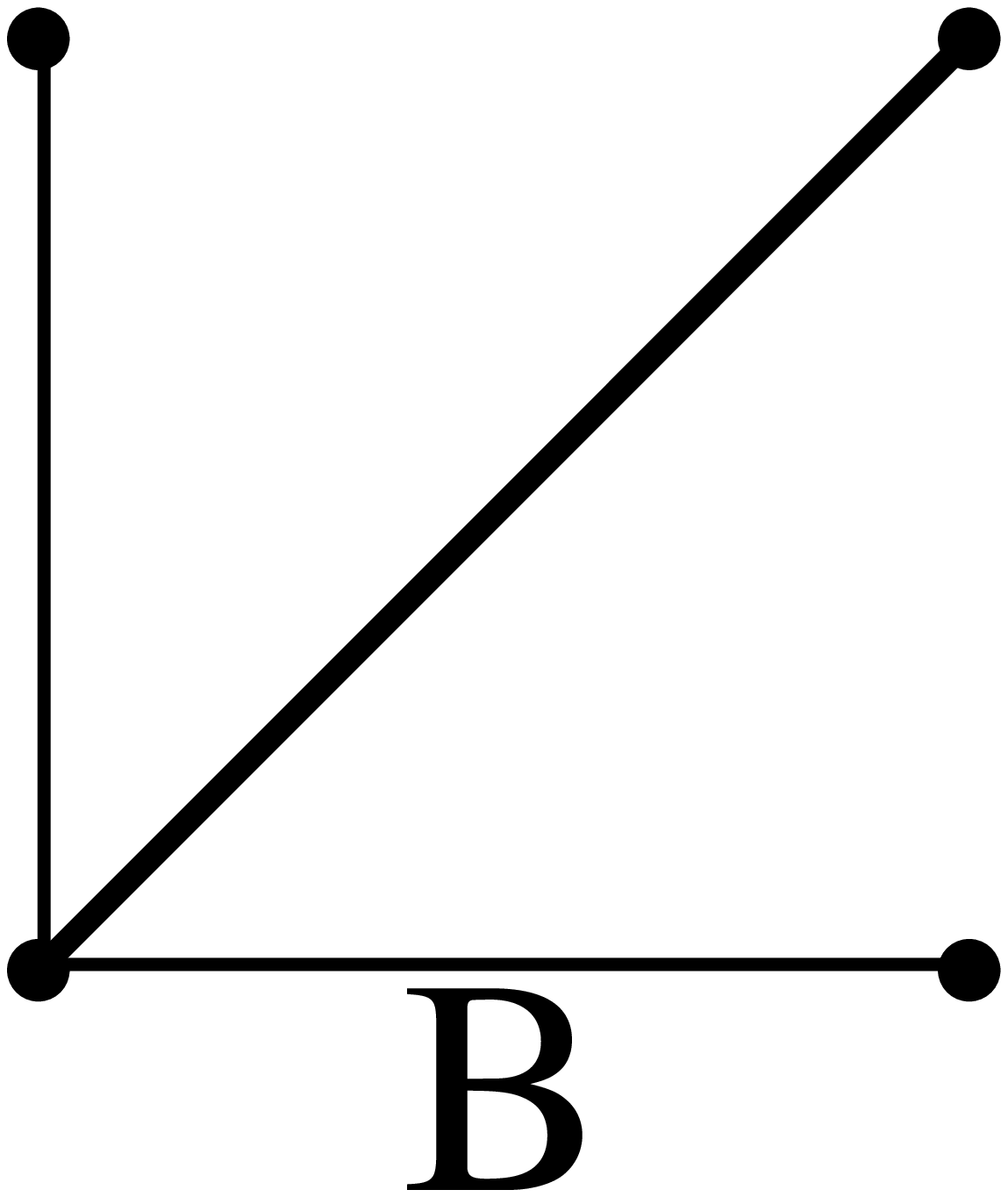}
\includegraphics[width=1.3cm]{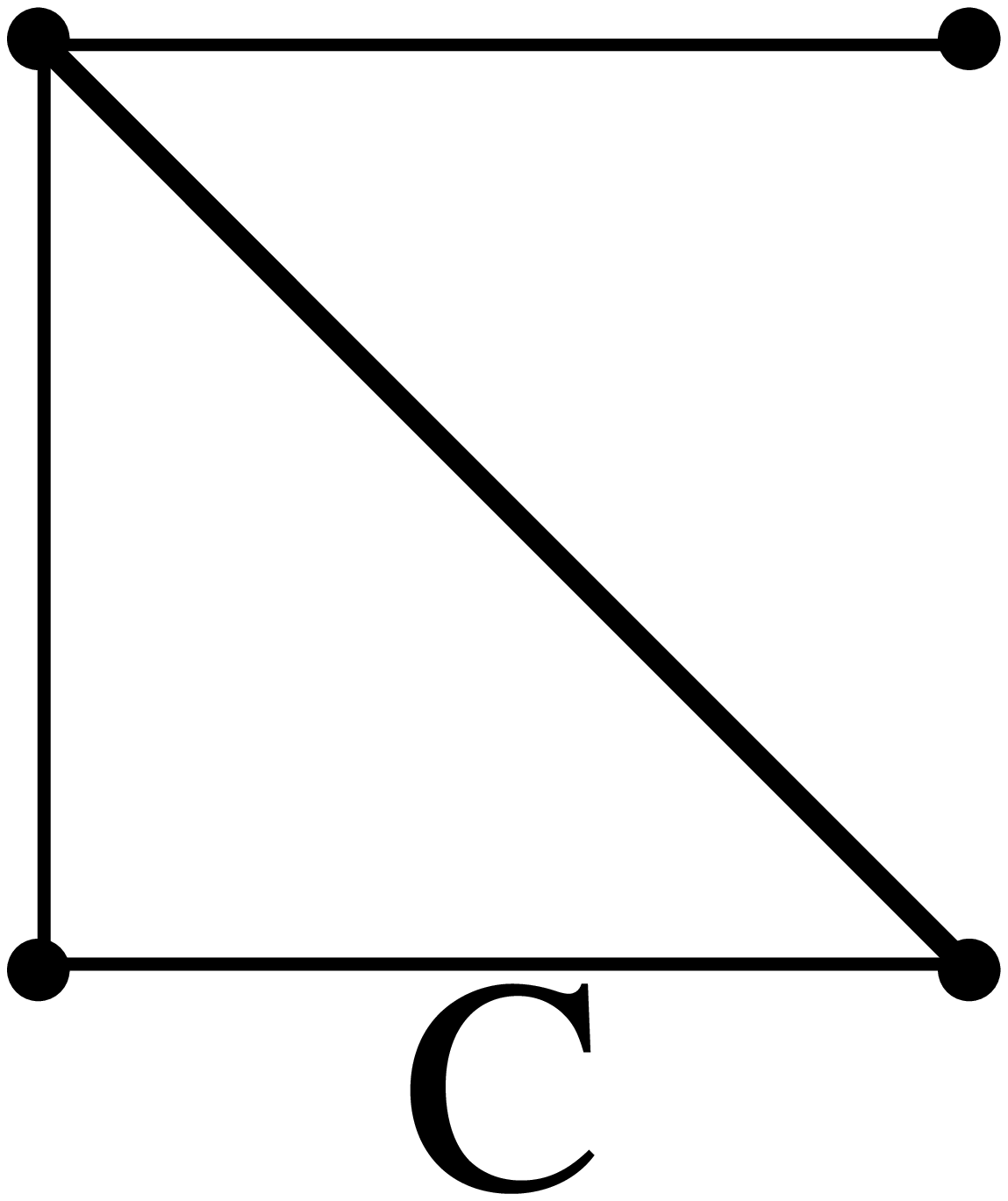}
\includegraphics[width=1.3cm]{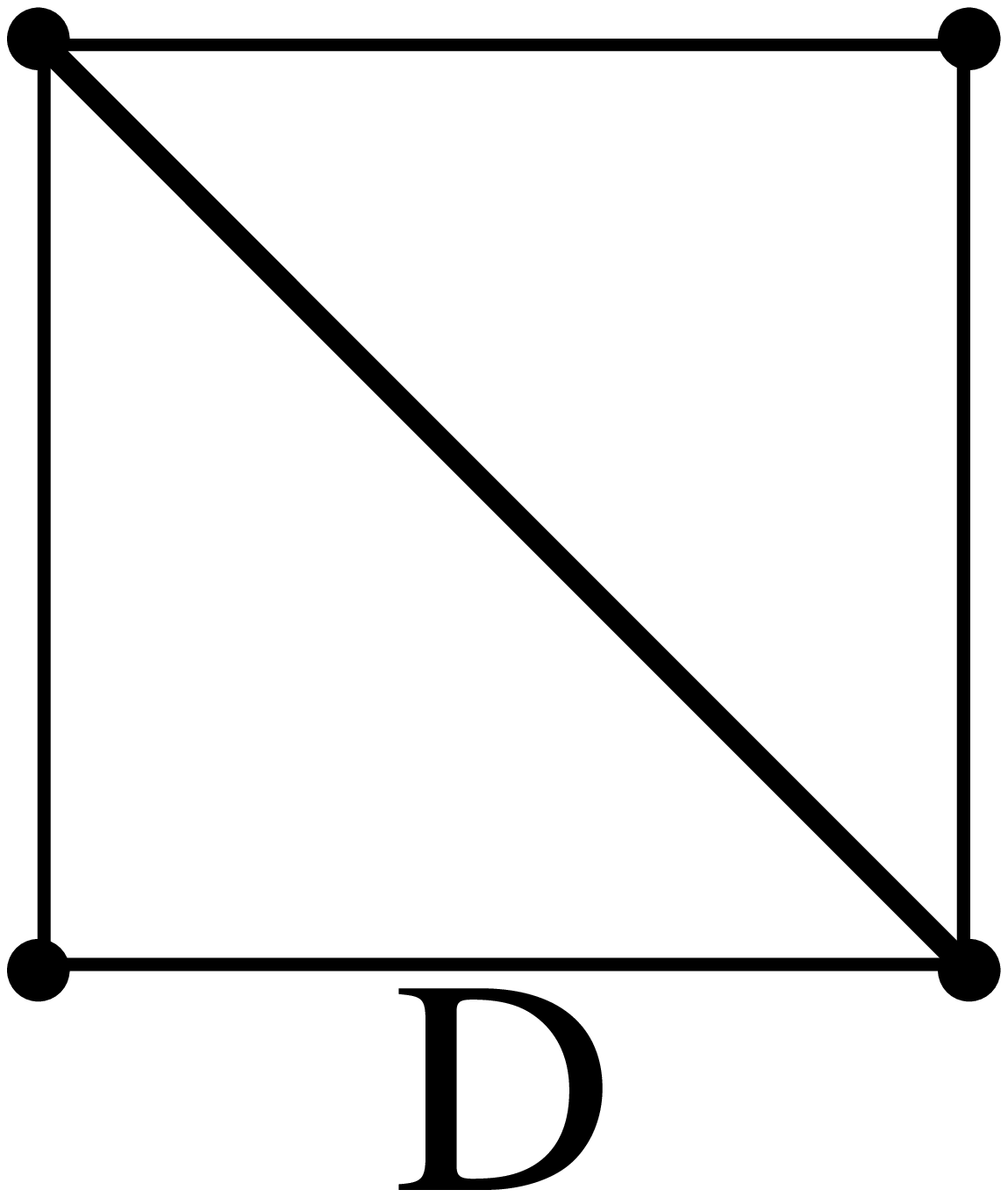}
\includegraphics[width=1.3cm]{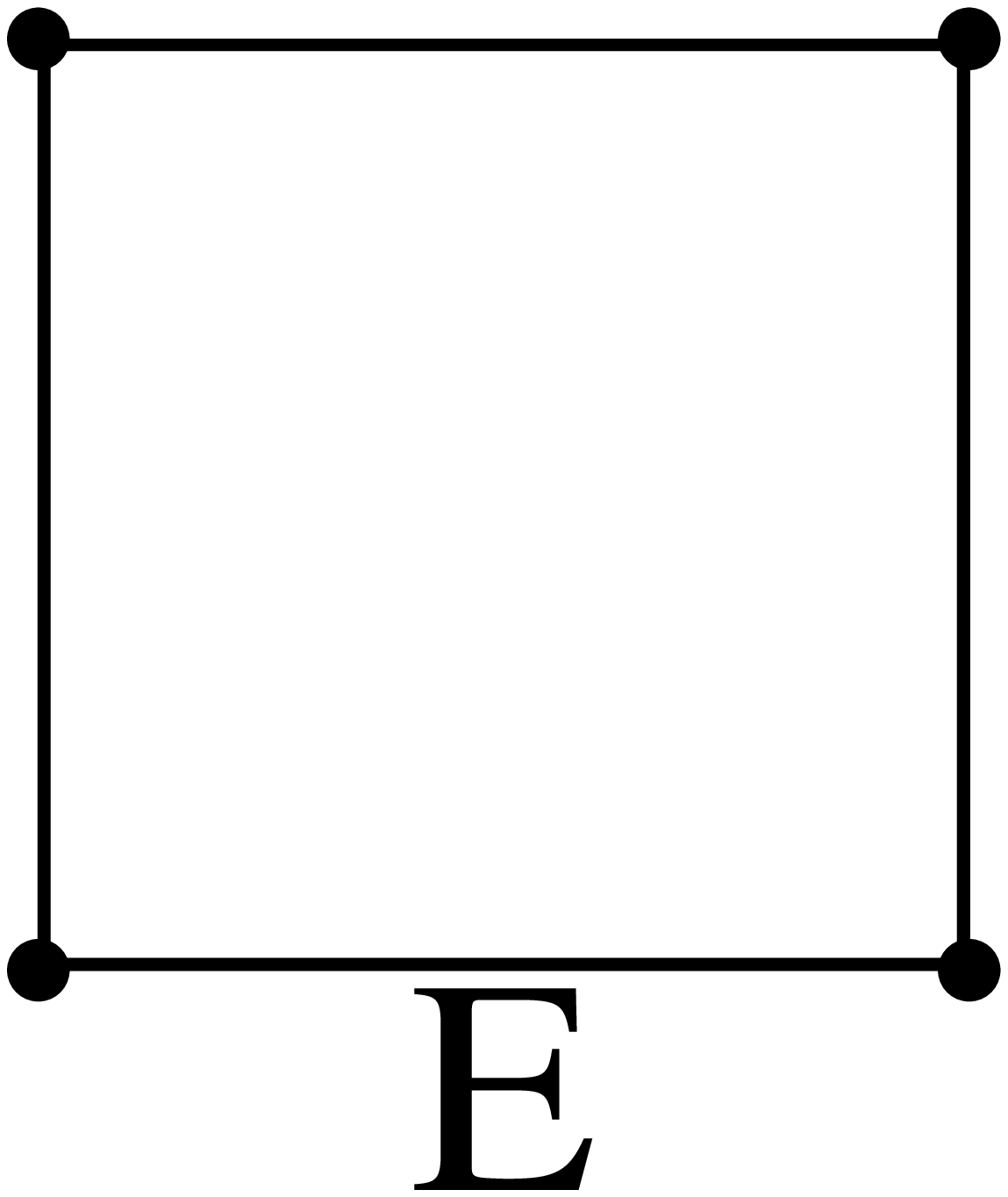}
\includegraphics[width=1.3cm]{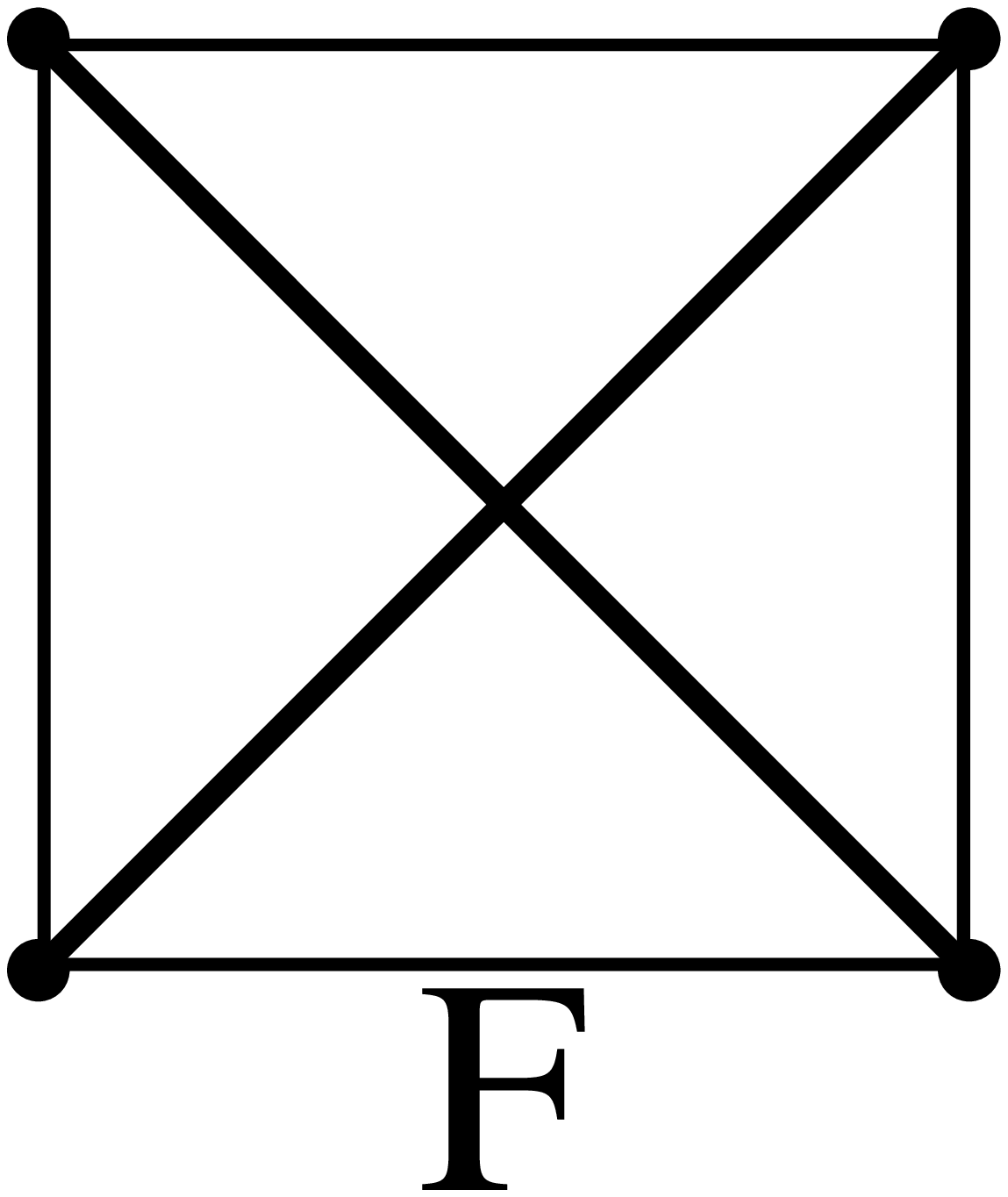}
\caption{\label{Fig:subgraph} All six network motifs of size 4  in undirected nearest-neighbor networks. These motifs are labeled A, B, C, D, E, and F, respectively.}
\end{figure}

Following Ref.~\cite{Xu-Zhang-Small1-2008-PNAS}, we consider motifs with four nodes. Fig.~\ref{Fig:subgraph} shows all possible motifs that are connected undirected graphs containing 4  four nodes. In general, different networks have different motif distributions, and it is not necessary that all motifs appear. For instance, a chain-like network or a loop has only motif A, a star-like graph has only motif B, and a complete graph has only motif F.
It is argued that distinct systems with the same motif occurrence rank may share some similar key elements  and  belong to a same superfamily \cite{Milo-Itzkovitz-Kashtan-Levitt-ShenOrr-Ayzenshtat-Sheffer-Alon-2004-Science,Xu-Zhang-Small1-2008-PNAS}.

\section{Numerical investigation of superfamily phenomenon in FBMs and MRWs}

\begin{figure}[htb]
\centering
\includegraphics[width=7cm]{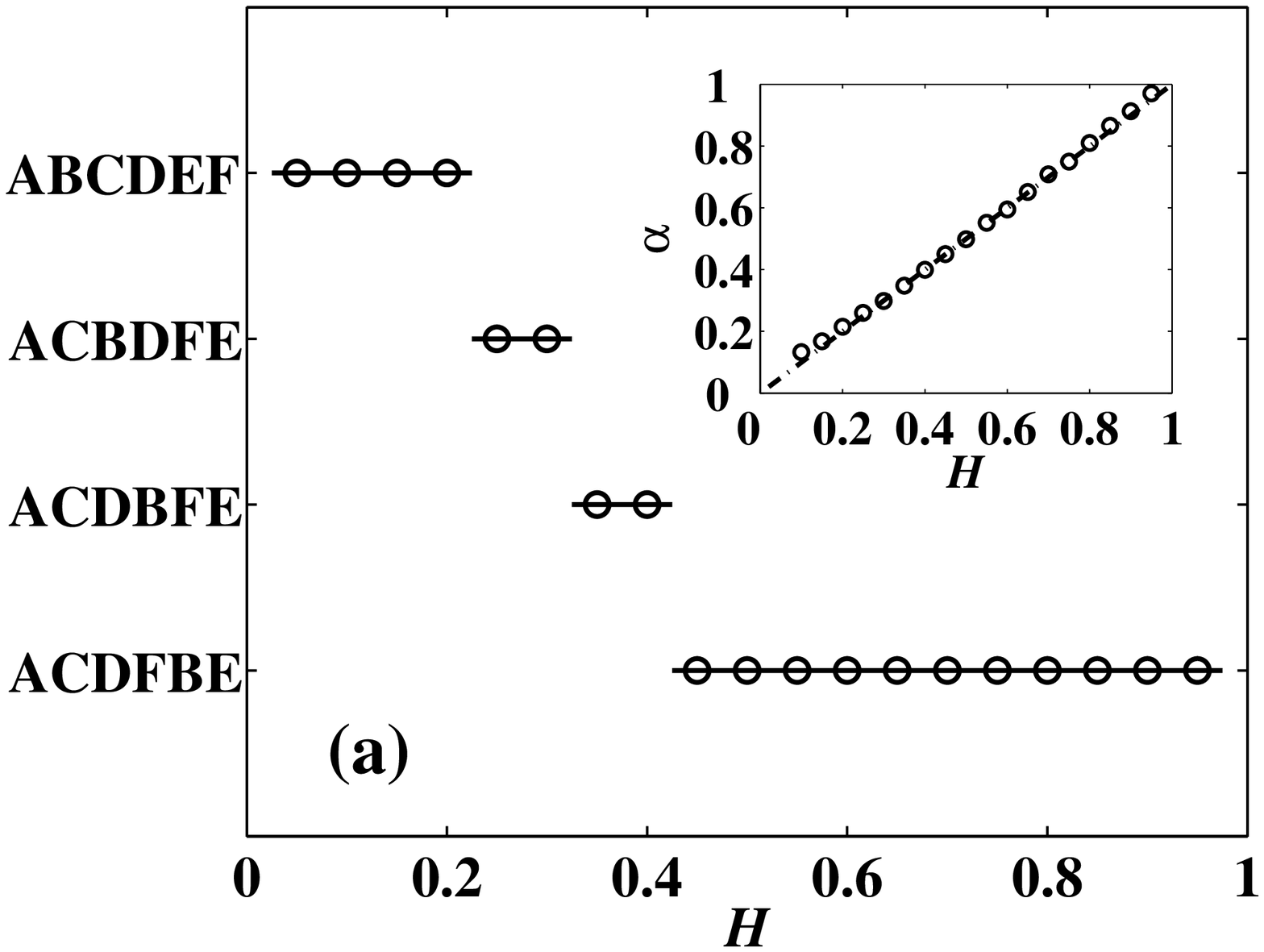}
\includegraphics[width=7cm]{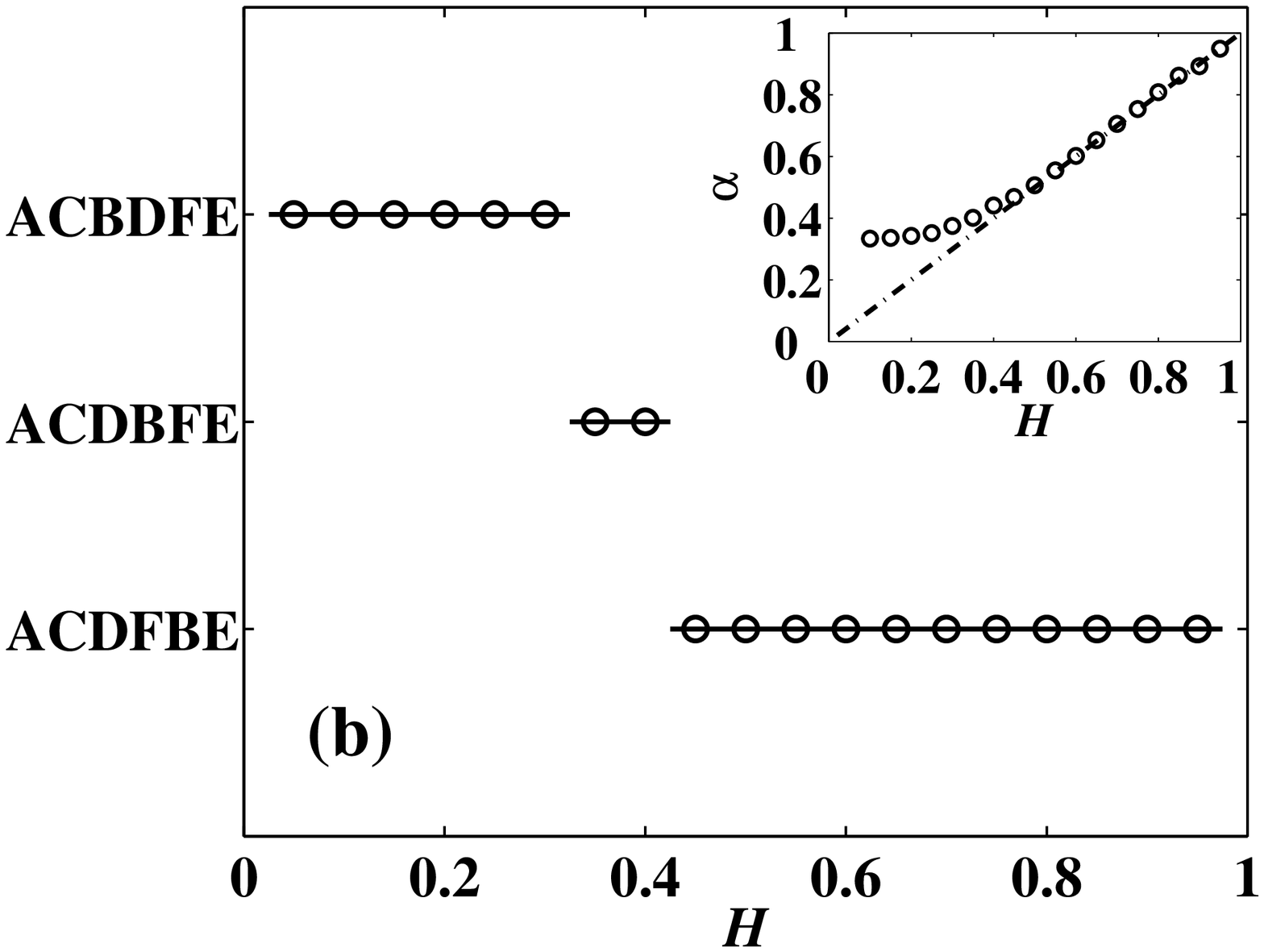}
\caption{\label{Fig:Motif:FBM:MRW} Network motif ranks of fractional Brownian motions (a) and multifractal random walks (b) with different Hurst indexes $H$. The two insets compare the DFA scaling exponent $\alpha$ of the sample series and the input Hurst index $H$ for the FBMs and MRWs based on 100 realizations.}
\end{figure}

We synthesize FBMs using wavelet-based algorithm to simulate FBMs \cite{Abry-Sellan-1996-ACHA} and MRWs \cite{Bacry-Delour-Muzy-2001-PRE}. For each input Hurst index $H$, which runs from 0.05 to 0.95 with a step of 0.05, 100 realizations of FBMs and MRWs are generated, and the corresponding motif distribution for each time series is obtained. Fig.~\ref{Fig:Motif:FBM:MRW} illustrates the dependence of the averaged motif ranks in descending order of the occurrence frequency with respect to the Hurst index $H$ for the two types of random walks. The insets of Fig.~\ref{Fig:Motif:FBM:MRW} compare the DFA scaling exponent $\alpha$ of the sample series and the input Hurst index $H$ for the FBMs and MRWs based on 100 realizations. Specifically, for each FBM (MRW) series with input Hurst index $H$, we calculate the DFA scaling exponent $\alpha$ based on the detrended fluctuation analysis \cite{Peng-Buldyrev-Havlin-Simons-Stanley-Goldberger-1994-PRE,Kantelhardt-Bunde-Rego-Havlin-Bunde-2001-PA}. It is found that $\alpha$ is very close to $H$ for the FBMs, while $\alpha$ is almost identical to $H$ when $H\geqslant0.45$ but systematically deviates from $H$ when $H<0.45$ for the MRWs \cite{Ni-Jiang-Zhou-2009-PLA}.

We observe that clear patterns emerge in Fig.~\ref{Fig:Motif:FBM:MRW}. For the FBMs, there are four motif rank patterns: ABCDEF when $H\leqslant0.20$, ACBDFE when $0.25\leqslant{H}\leqslant0.30$, ACDBFE when $0.35\leqslant{H}\leqslant0.40$, and ACDFBE when $0.45\leqslant{H}$. For the MRWs, there are three motif rank patterns: ACBDFE when $H\leqslant0.30$, ACDBFE when $0.35\leqslant{H}\leqslant0.40$, and ACDFBE when $0.45\leqslant{H}$. Strikingly, when we compare the two cases using the DFA scaling exponent $\alpha$, the two plots tell the same story: The superfamily classification is determined uniquely by the DFA scaling exponent. To be more clear, for $H\leqslant0.30$ in the case of MRWs in Fig.~\ref{Fig:Motif:FBM:MRW}(b), $\alpha$ is less than 0.35 and greater than 0.25, which corresponds to the motif rank pattern ACBDFE in Fig.~\ref{Fig:Motif:FBM:MRW}(a). It is not surprising that the pattern ABCDEF is missing in the case of MRWs, since it is hard to obtain MRW series with the DFA scaling exponent $\alpha$ smaller than 0.20. We can thus focus on the FBM case in the following discussion.

In Fig.~\ref{Fig:Motif:FBM:MRW}, we identify two key motifs, B and F, whose locations determine the associated motif rank patterns. If we ignore B and F, the four patterns reduce to the same combination ACDE. This phenomenon is also observed in all the cases presented in Ref.~\cite{Xu-Zhang-Small1-2008-PNAS}. The relative occurrence frequency (or rank) of motif B decreases with the increase of the DFA scaling exponent, and the rank of motif F increases with the DFA scaling exponent. We also find that, with the increase of the the DFA scaling exponent, the ranks of C and D increase and the rank of E decreases. These trends can be quantitatively explained as follows, especially for B and F.

Motif B occurs when the neighbors of the point are all far from each other, while motif F occurs when the four points are all close enough to each other. Therefore, motif B represents the most irregular structure while motif F indicates the most transitive condition. For large DFA scaling exponent, the time series is relatively more smooth. The consequence is that data points with close time moments in the time series are mapped to close points in the phase space that are more probable to connect with each other. More F motifs is expected to appear. For small DFA scaling exponent, the time series is relatively more rough. It follows that close points in the time series are mapped to a sparse region in phase space. In the sparse region, nontransitive structures such as motif B will be common \cite{Xu-Zhang-Small1-2008-PNAS}.

\section{Superfamily phenomenon in real systems}

We have shown above that correlated time series have a unique superfamily with a specific motif rank pattern ACDFBE while anticorrelated time series may exhibit four different motif rank patterns based on the associated DFA scaling exponents. In this section, we investigate the superfamily phenomenon in two real systems using stock market indexes and turbulence velocity. The similarity between stock markets and turbulence has been extensively studied \cite{Ghashghaie-Breymann-Peinke-Talkner-Dodge-1996-Nature,Mantegna-Stanley-1996-Nature}, which has stimulated the development of Econophysics \cite{Mantegna-Stanley-2000}. This comparative study aims to check if these two systems belong to the same superfamily.

\subsection{Example 1: Turbulence velocity signals}

The velocity data set in three-dimensional fully developed turbulence have been obtained from a hot-wire probe in the S1 wind tunnel of ONERA by the Grenoble group from LEGI \cite{Anselmet-Gagne-Hopfinger-Antonia-1984-JFM}. The mean velocity of the main flow is $ \langle{v}\rangle = 20 \rm{m s^{-1}}$, and the root mean square velocity fluctuation is $v_{\rm{rms}} = 1.684 \rm{ms^{-1}}$. Thus the turbulence intensity is calculated to be $I = {v_{\mathtt{rms}}} / {\langle{v}\rangle} = 0.0842$ and the velocity is large enough compared with turbulent fluctuations to use the Taylor's frozen flow hypothesis. The integral scale $\approx 2 \rm{m}$ is obtained from the autocorrelation of the velocity series (according to the Taylor's frozen flow hypothesis). The Kolmogorov microscale $\eta$ is 0.195 mm, the Taylor microscale $\lambda =16.6$ mm, and the Taylor-scale Reynolds number is ${\rm{Re}}_\lambda = 1863$ \cite{Zhou-Sornette-2002-PD}.

\begin{figure}[htb]
  \centering
  \hspace{6mm}\includegraphics[width=7.4cm]{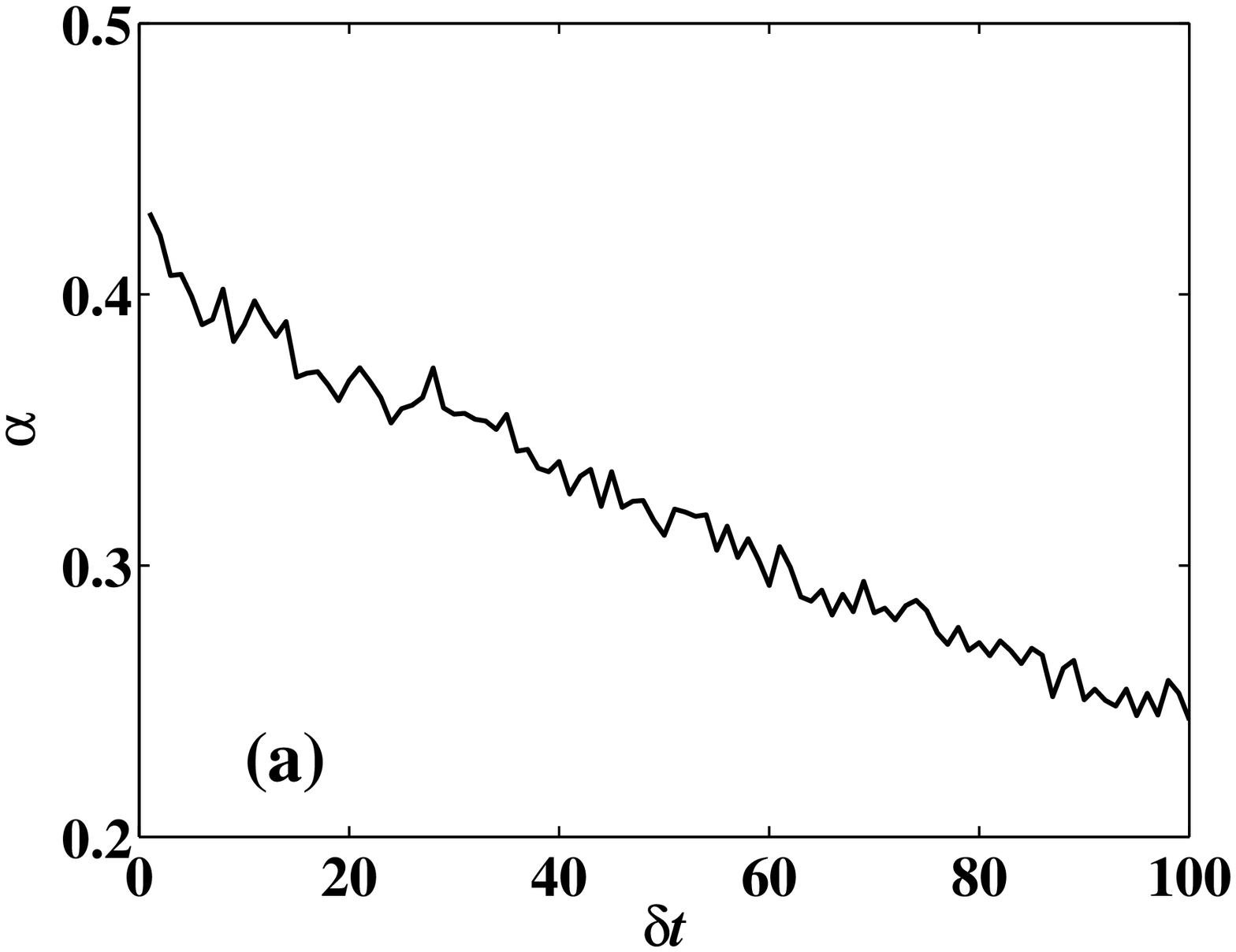}
            \includegraphics[width=8cm]{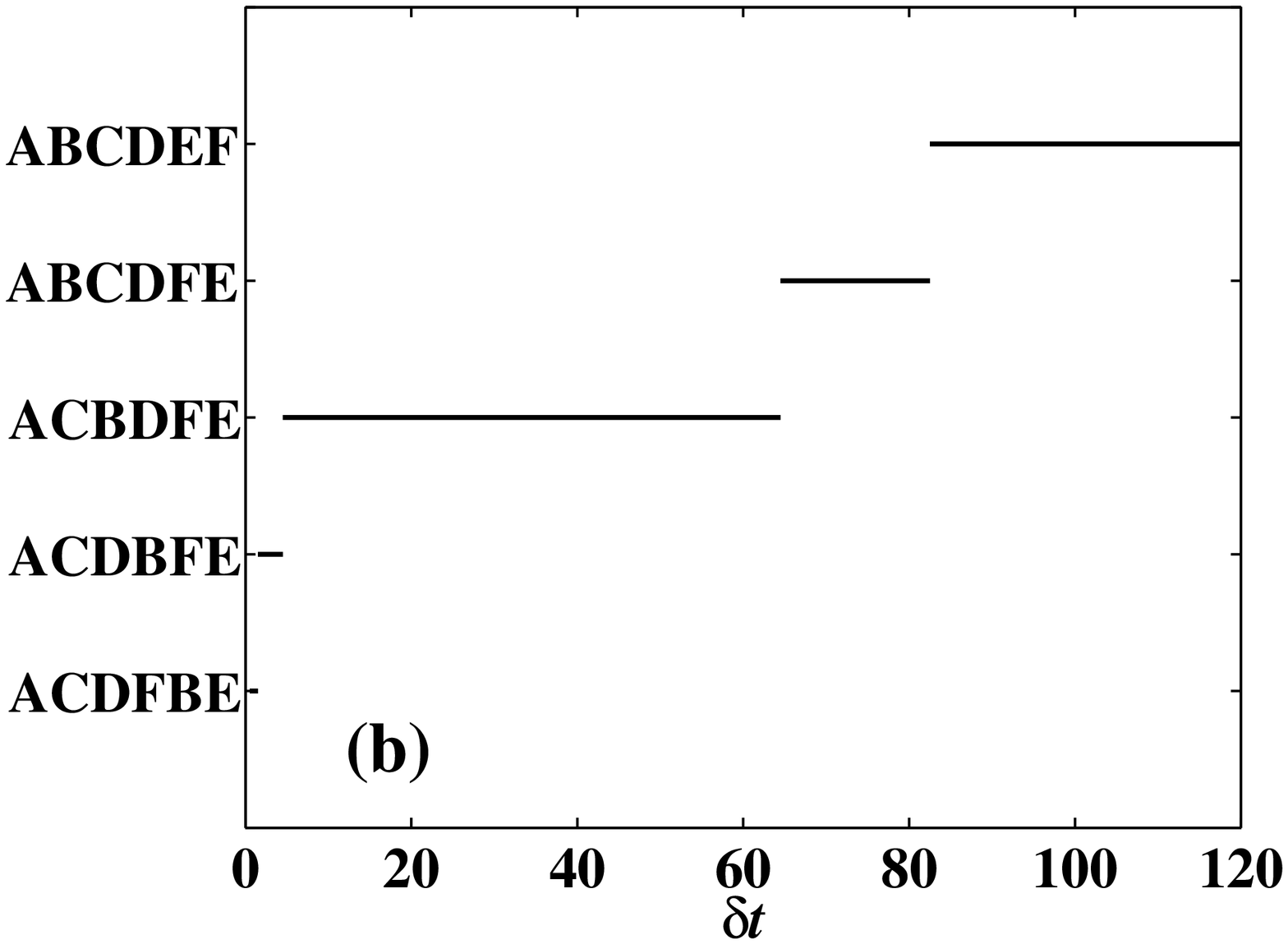}
  \caption{\label{Fig:Turbulence:Hurst:Motif}(a) Dependence of the DFA scaling exponent $\alpha$ of the resampled turbulence signals with respect to the time scale $\delta{t}$. (b) Dependence of the motif rank pattern on the resampling interval $\delta{t}$. The values with $\delta{t}$ larger than 120 are not shown.}
\end{figure}

The measurement resolution of the experiment is $3.55\times10^{-5}$s. The total length of the velocity signal is about $L=1.7\times10^7$. The signal is sampled at different time scales $\delta{t}$ (in units of the measurement resolution). In our analysis, $\delta{t}$ ranges from 1 to 200 with an increment of 1. For each $\delta{t}$, the length of the sampled signal is about $L_{\delta{t}}=L/\delta{t}$, and 100 randomly subseries with length $\ell=10^4$ are extracted for analysis. The estimated DFA scaling exponents $\alpha$ are shown in Fig.~\ref{Fig:Turbulence:Hurst:Motif}(a) as a function of the sampling time scale $\delta{t}$, and the motif rank patterns for each $\delta{t}$ are illustrated in Fig.~\ref{Fig:Turbulence:Hurst:Motif}(b).

\begin{figure*}[htb]
  \centering
  \includegraphics[width=5cm]{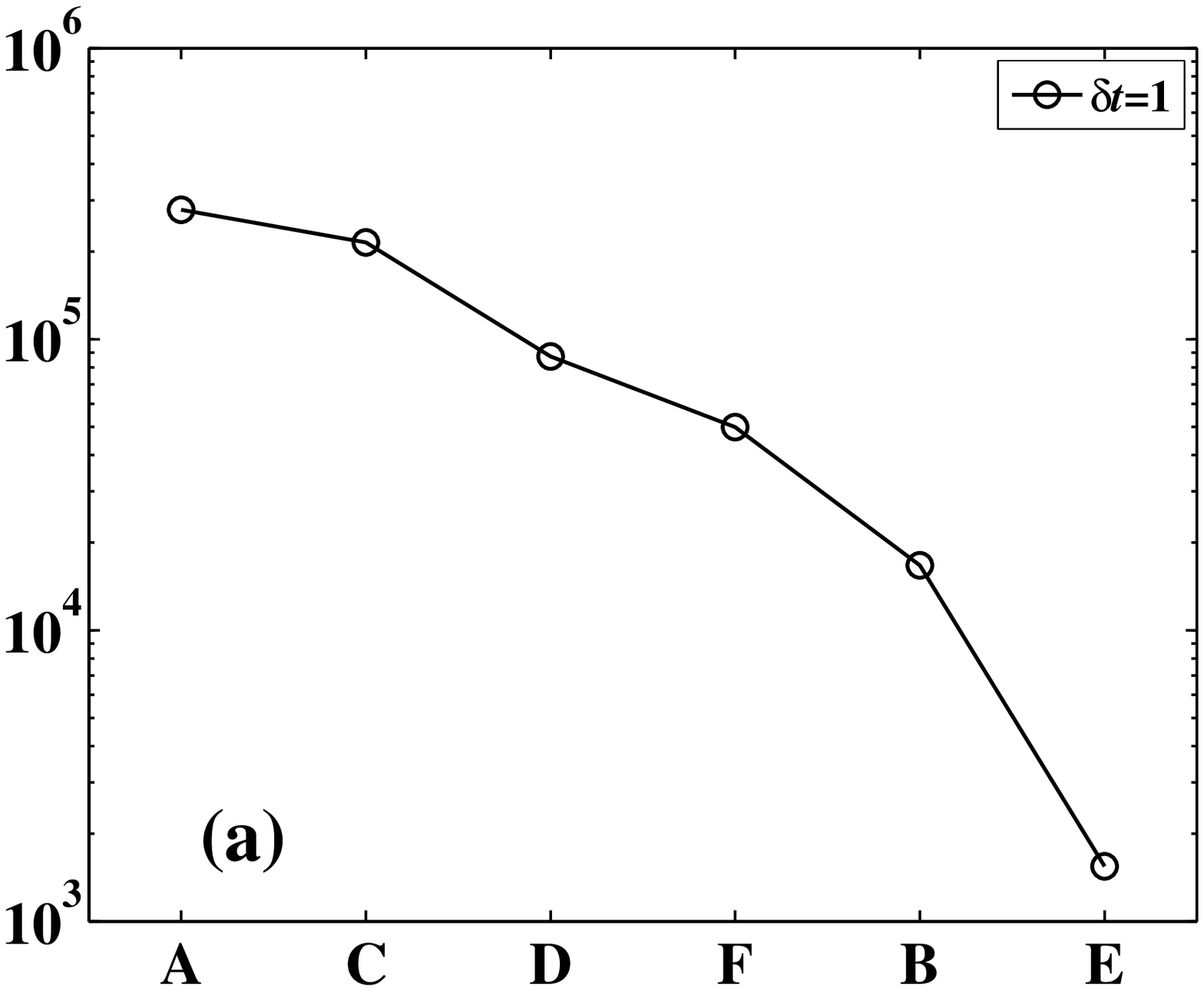}
  \includegraphics[width=5cm]{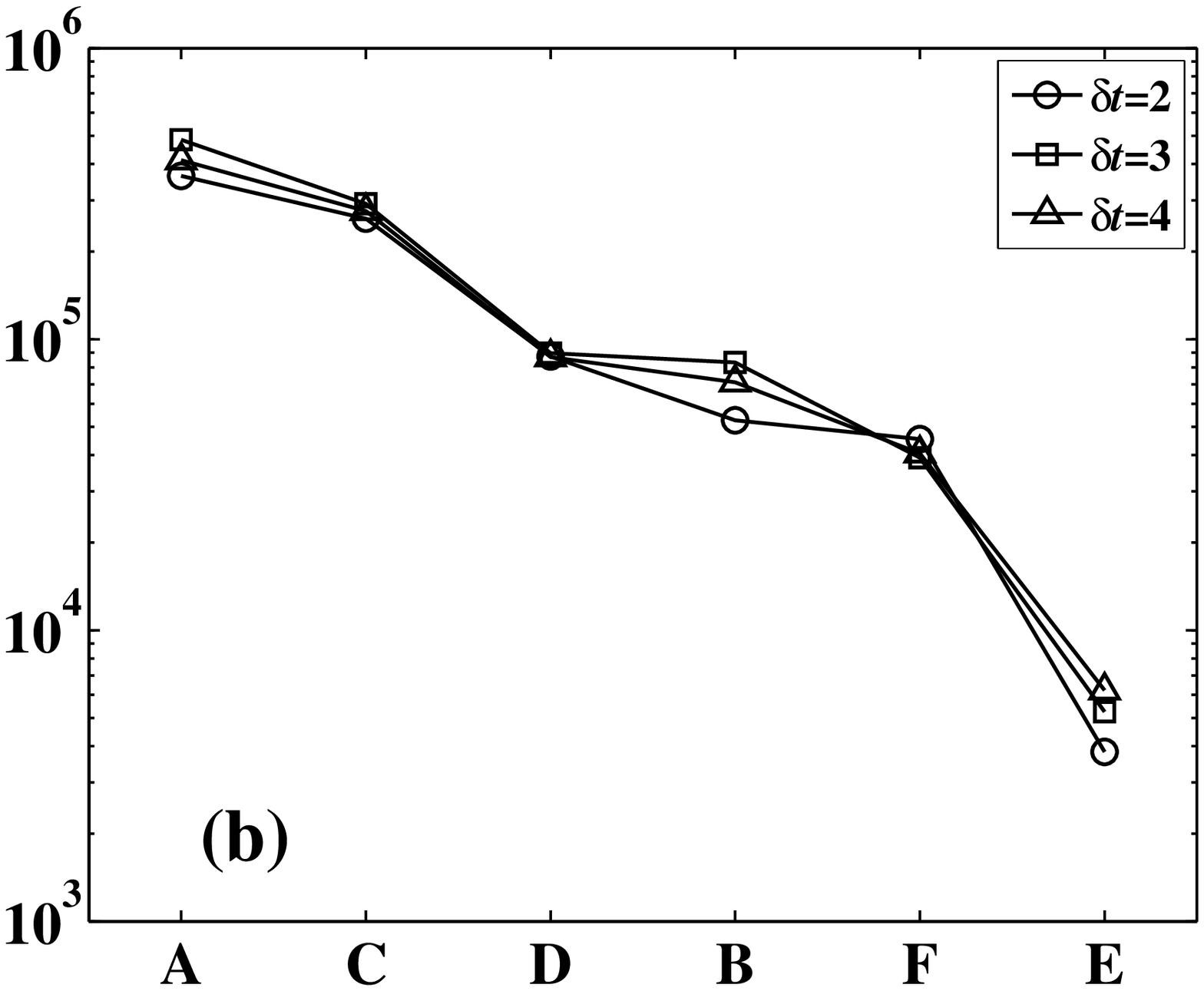}
  \includegraphics[width=5cm]{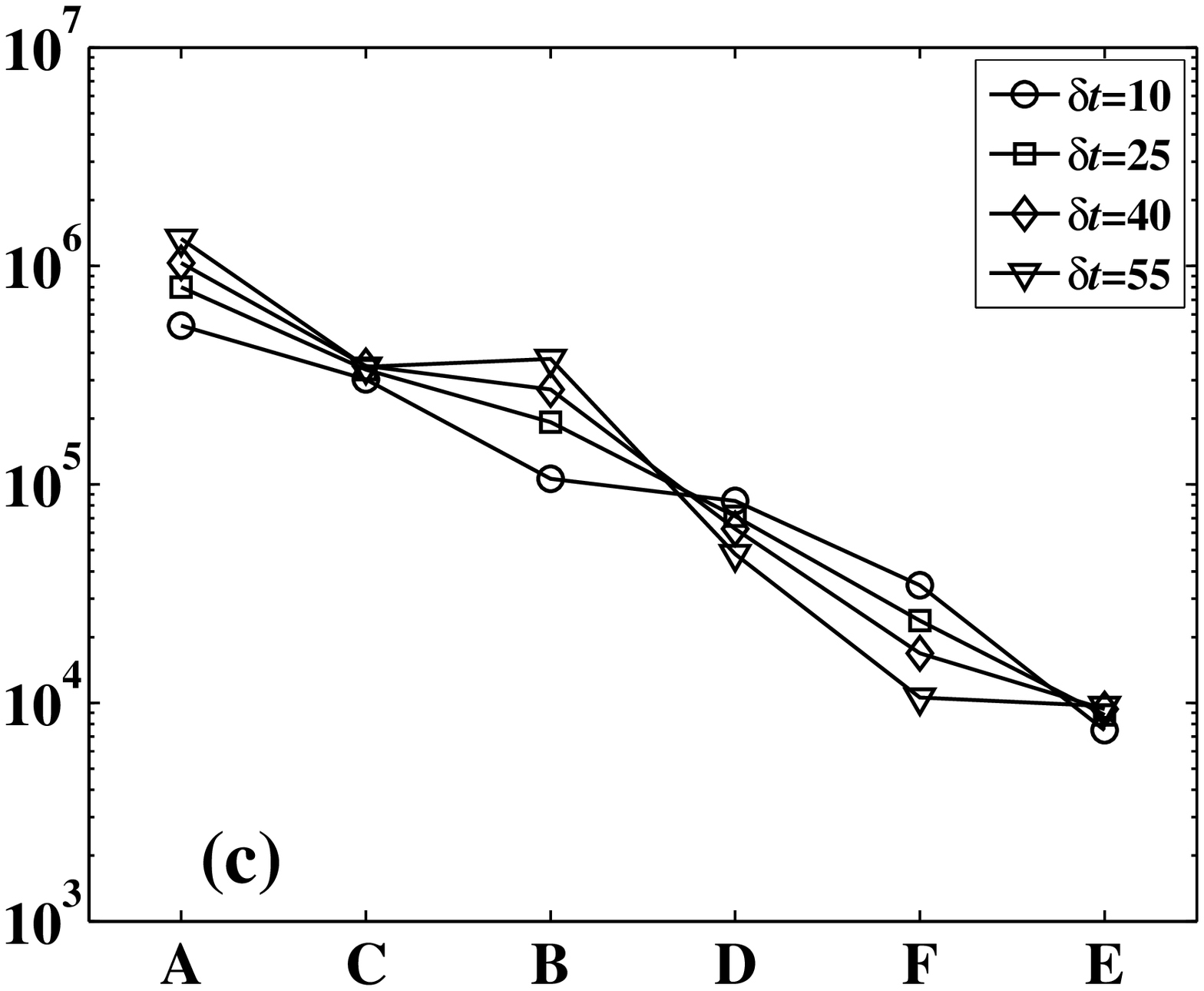}
  \includegraphics[width=5cm]{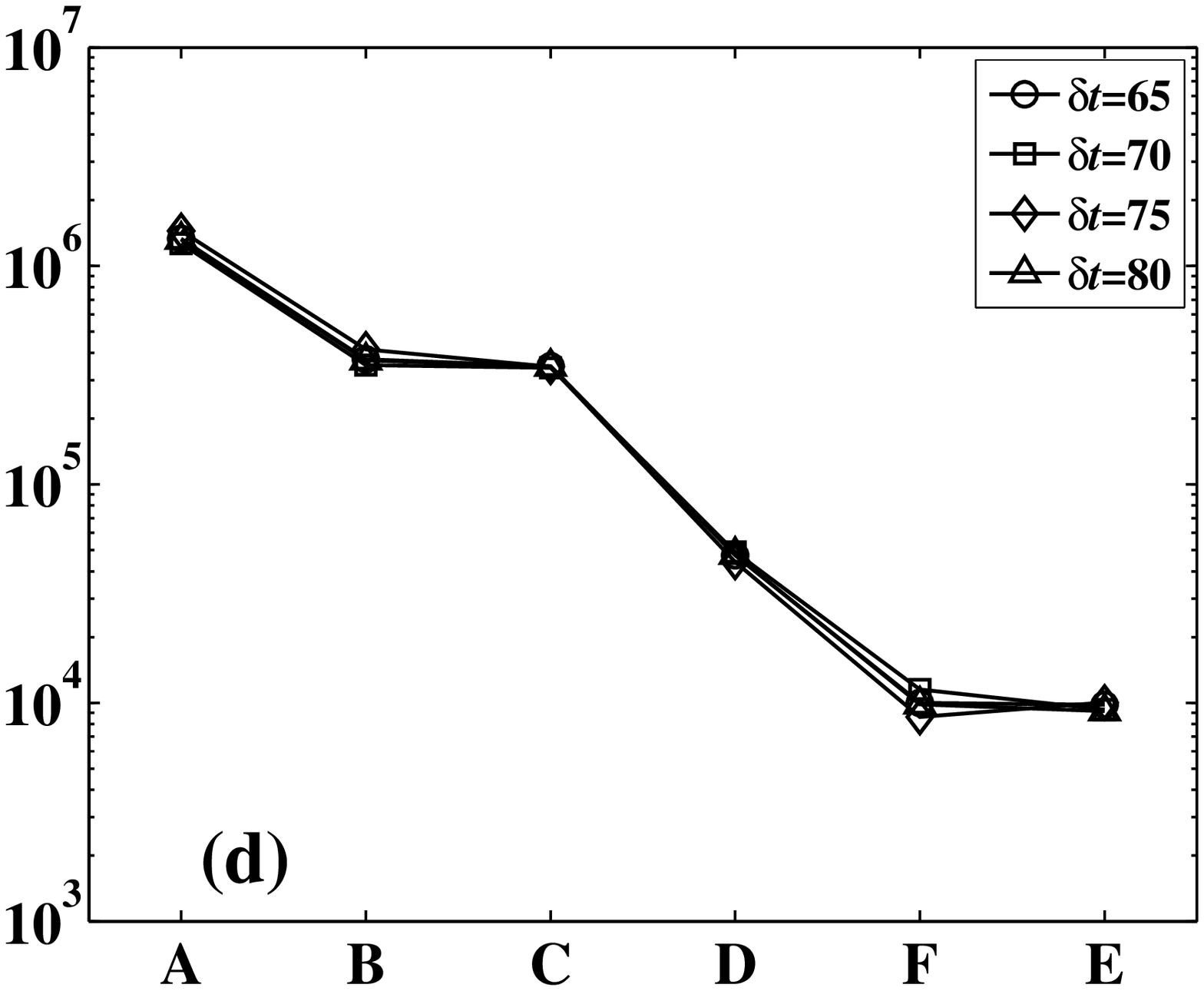}
  \includegraphics[width=5cm]{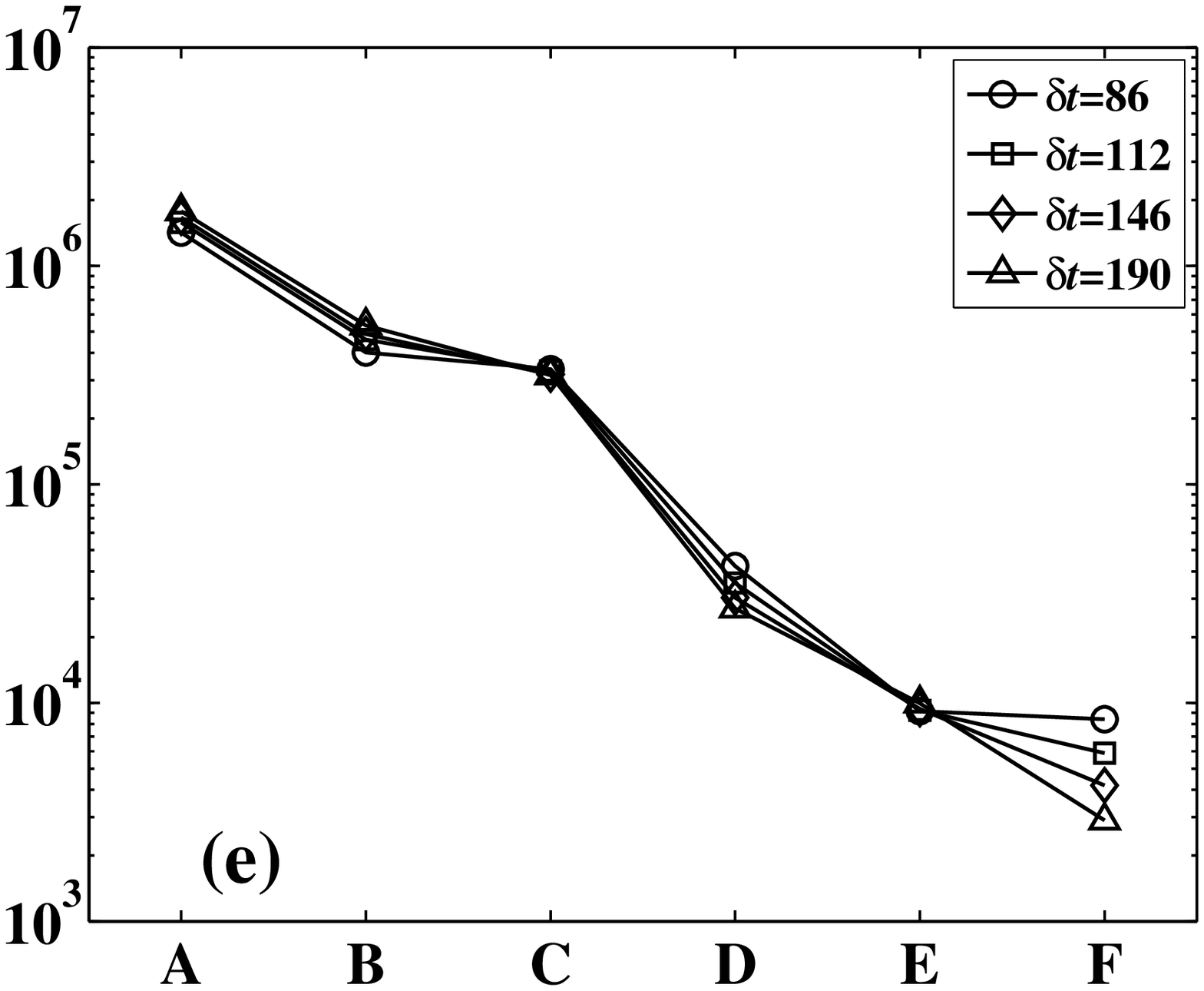}
  \includegraphics[width=5cm,height=4.1cm]{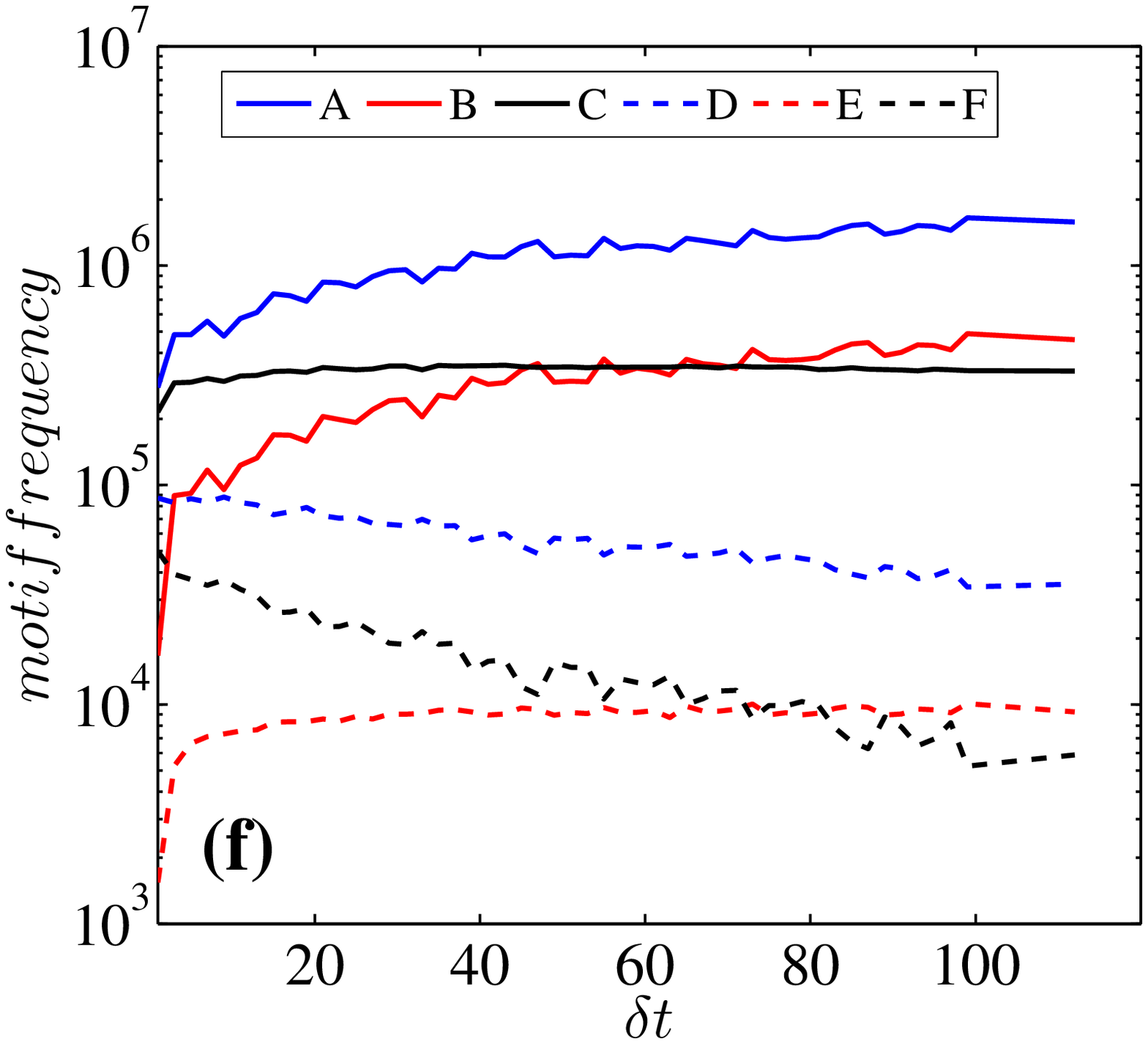}
  \caption{\label{Fig:Turbulence:MotifRank} Network motif ranks of turbulence signals with different sampling time scales: (a) $\delta{t}=1$, (b) $2\leqslant\delta{t}\leqslant4$, (c) $5\leqslant\delta{t}\leqslant64$, (d) $65\leqslant\delta{t}\leqslant82$, and (e) $83\leqslant\delta{t}$.}
\end{figure*}

In Fig.~\ref{Fig:Turbulence:Hurst:Motif}(b), we observe five motif rank orders: ACDFBE for $\delta{t}=1$, ACDBFE for $2\leqslant\delta{t}\leqslant4$, ACBDFE for $5\leqslant\delta{t}\leqslant64$, ABCDFE for $65\leqslant\delta{t}\leqslant82$, and ABCDEF for $83\leqslant\delta{t}$. Taking into consideration the fact that the motif rank pattern ABCDFE is very close to ABCDEF as we will show below, it is found that, with the decrease of $\delta{t}$ (thus the increase of $\alpha$), the motif rank pattern changes in an order of ABCDEF, ACBDFE, ACDBFE, and ACDFBE. Furthermore, the switching values of $\alpha$ from one pattern to the other are also in excellent agreement with Fig.~\ref{Fig:Motif:FBM:MRW}. Fig.~\ref{Fig:Turbulence:Hurst:Motif} also implies that the turbulence velocity signal possesses multiscale nature. However, the switching values of $\delta{t}$ cannot be related to the well-known turbulence scales (such as the dissipation subrange and the inertial subrange).

Plots (a-e) of Fig.~\ref{Fig:Turbulence:MotifRank} illustrate typical motif ranks in descending order for the networks induced from velocity series at different time scales. Fig.~\ref{Fig:Turbulence:MotifRank}(c) shows that the occurrence frequency of motif B (and motif A as well) decreases with $\alpha$ and the frequency of motif F (and D as well) increases with $\alpha$, as explained for the FBM case. Fig.~\ref{Fig:Turbulence:MotifRank}(e) confirms this observation since the occurrence frequency of motif F decreases with $\delta{t}$ and thus increases with $\alpha$. There are also time series (say, $\delta{t}=55$ in Fig.~\ref{Fig:Turbulence:MotifRank}(c) and $\delta{t}=75$ in Fig.~\ref{Fig:Turbulence:MotifRank}(d)) that have different motif rank patterns with their ``neighbors''. However, this situation is rare, and averaging over 100 realizations with the same $\delta{t}$ gives consistent and stable results. In addition, we find that the relative frequencies of motif E and F are comparable in Fig.~\ref{Fig:Turbulence:MotifRank}(d), which makes the motif rank pattern ABCDFE marginal when compared with the pattern ABCDEF.

The above observations are further illustrated in Fig.~\ref{Fig:Turbulence:MotifRank}(f). We find that, with the increase of $\delta{t}$, the motif frequencies of A, B, C and E increase, while the motif frequencies of D and F decrease. There are two decreasing quaternionic combinations, ACDE and ACDF. In addition, there are four intersections caused by B and C, B and E, B and F, and E and F, which correspond to the four switching values of $\delta{t}$ separating the five motif rank patterns shown in Fig.~\ref{Fig:Turbulence:MotifRank}(a-e).

\subsection{Example 2: Stock market indexes}

\begin{figure}[htb]
  \centering
  \includegraphics[width=8cm]{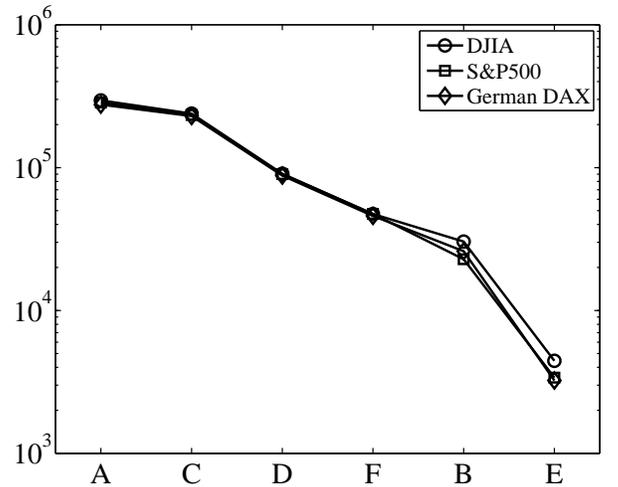}
  \caption{\label{Fig:Stock:MotifRank} Network motif ranks of the nearest-neighbor networks induced from three stock market indexes.}
\end{figure}

Now we study the superfamily phenomenon of three stock market indexes: The DJIA from 26-May-1896 to 27-Apr-2007 of size 30147, the S\&P 500 index from 29-Nov-1940 to 07-Jan-2008 of size 16943, and the German DAX index from 01-Oct-1959 to 15-Nov-2002 of size 10812. For these three data sets, it is hard to perform a similar analysis as for the turbulence signal due to the short length of the data. Therefore, we do not perform multiscale analysis on the stock index data. We convert the time series into three nearest-neighbor networks in the phase space and count the occurrence of the six motifs. The motif ranks for the three stock indexes are shown in Fig.~\ref{Fig:Stock:MotifRank}.

According to Fig.~\ref{Fig:Stock:MotifRank}, the three stock indexes have the same motif ranks ACDFBE. This is consistent with the regime of $\alpha\geqslant0.45$ in the fractional Brownian motion case illustrated in Fig.~\ref{Fig:Motif:FBM:MRW}(a). It is well-known that stock index returns are basically uncorrelated with the DFA scaling exponent $\alpha\approx0.5$ \cite{Mantegna-Stanley-2000}. It is evident that the superfamily phenomenon of stock indexes confirms the main conclusion of this Letter.

\section{Conclusion}

Time series can be mapped into nearest-neighbor networks, which allows us to classify time series into different superfamilies based on the motif occurrence patterns. We have investigated the motif patterns of fractional Brownian motions and multifractal random walks and found that the DFA scaling exponent of the time series plays a unique role in the superfamily classification. The multifractal nature of the MRWs has no influence on the motif ranks. Four different motif rank patterns ABCDEF, ACBDFE, ACDBFE, and ACDFBE are identified, which correspond to different ranges of the DFA scaling exponent. The results of similar analyses of velocity time series in fully developed turbulence and daily stock market indexes are essentially consistent with this finding, except that an additional pattern ABCDEF appears between ABCDEF and ACBDFE with a very narrow range of the exponent $\alpha$ for the turbulence case.

{\textbf{Acknowledgments:}}

This work was partially supported by the Program for Changjiang Scholars and Innovative Research Team in University (IRT0620) and the Program for New Century Excellent Talents in University (NCET-07-0288).

\bibliography{E:/Papers/Auxiliary/Bibliography}


\end{document}